\documentclass[letter]{IEEEtran}
\usepackage{graphicx}
\usepackage{fancyhdr}
\usepackage{makeidx}
\usepackage{amssymb,amsmath}
\usepackage{enumerate}
\newcommand{\bd}{\begin{document}}
\newcommand{\ed}{\end{document}}
\newcommand{\bc}{\begin{center}}
\newcommand{\ec}{\end{center}}
\newcommand{\vs}{\vspace}
\newcommand{\hs}{\hspace}
\newcommand{\beq}{\begin{equation}}
\newcommand{\eeq}{\end{equation}}
\newcommand{\beqs}{\begin{eqn*}}
\newcommand{\eeqs}{\end{eqn*}}
\newcommand{\bq}{\begin{quote}}
\newcommand{\eq}{\end{quote}}
\newcommand{\lb}{\linebreak}
\newcommand{\mb}{\makebox}
\newcommand{\fb}{\framebox}
\newcommand{\mc}{\multicolumn}
\newcommand{\ben}{\begin{enumerate}}
\newcommand{\een}{\end{enumerate}}
\newcommand{\bit}{\begin{itemize}}
\newcommand{\eit}{\end{itemize}}
\newcommand{\ov}{\overline}
\newcommand{\un}{\underline}
\newcommand{\lt}{\left}
\newcommand{\rt}{\right}
\newcommand{\ba}{\begin{array}}
\newcommand{\ea}{\end{array}}
\newcommand{\beqa}{\begin{eqnarray}}
\newcommand{\eeqa}{\end{eqnarray}}
\newcommand{\beqas}{\begin{eqnarray*}}
\newcommand{\eeqas}{\end{eqnarray*}}
\newcommand{\bfg}{\begin{figure}}
\newcommand{\efg}{\end{figure}}
\newcommand{\pad}{\partial}
\newcommand{\nn}{\nonumber}
\newcommand{\la}{\leftarrow}
\newcommand{\ra}{\rightarrow}
\newcommand{\lgla}{\longleftarrow}
\newcommand{\lgra}{\longrightarrow}
\newcommand{\La}{\Leftarrow}
\newcommand{\Ra}{\Rightarrow}
\newcommand{\Lra}{\Leftrightarrow}
\newcommand{\Lgla}{\Longleftarrow}
\newcommand{\Lgra}{\Longrightarrow}
\renewcommand{\a}{\alpha}
\renewcommand{\b}{\beta}
\newcommand{\g}{\gamma}
\newcommand{\G}{\Gamma}
\renewcommand{\d}{\delta}
\newcommand{\D}{\Delta}
\newcommand{\e}{\epsilon}
\newcommand{\eps}{\epsilon}
\newcommand{\s}{\sigma}
\renewcommand{\l}{\lamda}
\newcommand{\m}{\mu}
\newcommand{\n}{\nu}
\renewcommand{\S}{\Sigma}
\newcommand{\p}{\pi}
\newcommand{\om}{\omega}
\newcommand{\Om}{\Omega}
\newcommand{\tri}{\triangle}
\newcommand{\ti}{\times}
\newcommand{\f}{\frac}
\newcommand{\ds}{\displaystyle}
\newcommand{\bm}[1]{\mb{{\boldmath $#1$}}}
\newcommand{\alter}[2]{\lt\{ \ba{ll}#1 \\ #2 \ea \rt.}
\newcommand{\alt}[4]{\lt\{ \ba{ll}#1 & \mb{if \, \,}#2 \\ #3 & \mb{}#4 \ea
    \rt.}
\newcommand{\altn}[4]{\lt\{ \ba{rl}#1 & \mb{if \, \,}#2 \\ #3 & \mb{}#4 \ea
    \rt.}
\newcommand{\altif}[4]{\lt\{ \ba{ll}#1 & \mb{if \, \,}#2 \\ #3 &
\mb{if \, \,}#4 \ea \rt.}
\newcommand{\altnif}[4]{\lt\{ \ba{rl}#1 & \mb{if \, \,}#2 \\ #3 &
\mb{if \, \,}#4 \ea \rt.}
\newcounter{algc}
\newcounter{romc}
\newcounter{Alphc}
\newcommand{\bl}{\begin{list}{{\it Step} ~\arabic{algc}~:} {\usecounter{algc}
                \setlength{\topsep}{0pt} \setlength{\itemsep}{0pt}}}
\newcommand{\el}{\end{list}}
\newcommand{\blr}{\begin{list}{~\roman{romc}~:} {\usecounter{romc}
                \setlength{\topsep}{0pt} \setlength{\itemsep}{0pt}}}
\newcommand{\elr}{\end{list}}
\newcommand{\bla}{\begin{list}{~\Alph{Alphc}~:} {\usecounter{Alphc}
                \setlength{\topsep}{0pt} \setlength{\itemsep}{0pt}}}
\newcommand{\ela}{\end{list}}

\newtheorem{theorem}{Theorem}

\begin{document}
\bstctlcite{IEEEexample:BSTcontrol}
\title{HFinFET: A Scalable, High Performance, Low Leakage Hybrid N-Channel FET}
\author{Kausik Majumdar, Prashant Majhi, Navakanta Bhat and Raj Jammy
\thanks{K. Majumdar and N. Bhat are with the
Department of Electrical Communication Engineering and the Center of Excellence in Nanoelectronics,
Indian Institute of Science, Bangalore-560012, India. Email: kausik@ieee.org.

P. Majhi is Intel Assignee at the Sematech International, 2706 Montopolis Dr, Austin, TX-78741, US.

R. Jammy is with the Sematech International, 2706 Montopolis Dr,
Austin, TX-78741, US. }}
\date{}
\maketitle
{\abstract In this letter we propose the design and simulation study of a novel transistor,
called HFinFET, which is a hybrid of a HEMT and a FinFET, to obtain excellent performance
and good off state control. Followed by the description of the design, 3D device simulation
has been performed to predict the characteristics of the device. The device has been
benchmarked against published state of the art HEMT as well as planar and non-planar
Si NMOSFET data of comparable gate length using standard benchmarking techniques.}

{\keywords FinFET, HEMT, Transistor Scaling, Coupled Poisson-Schrodinger Equations.}
\section{Introduction}
Due to scalability issues of bulk MOSFET, beyond 22nm technology node, it is necessary to look for
alternative options to drive the semiconductor industry.
FinFET is one of the proposed candidates which shows excellent subthreshold slope and hence
scalability due to its nonplanar multi-gate structure \cite{ykc01,fly06,by02}. On the other hand,
HEMT devices \cite{k07,k08} and some variants of it \cite{d05} drive the performance
of ultra-fast devices using III-V channel material. However, their off state
control, gate leakage and scalability need to be addressed before using them in VLSI
logic circuit applications. Hence the bulk planar MOSFET still remains the major technology
driver in semiconductor industry.

The aim of this
work is to come up with a novel solution that takes care of both high performance
and low leakage, but staying in the quasi-MOS regime which makes it possible to
fabricate the device without extensive deviation from the existing technology. In the
rest of the paper, we propose a scalable hybrid device, called HFinFET,
to meet such requirements, namely
providing high ON state current using bulk conduction arising from HEMT-like
mechanism, and good off-state control by FinFET-like mechanism. Followed by
the description of the device in sec. \ref{sec:proposal}, we discuss the
characteristics of the device in sec. \ref{sec:resdults}. The four benchmarking
techniques proposed in \cite{c05}, namely (1) intrinsic gate delay ($\tau$) vs gate length ($L_g$),
(2) subthreshold slope (SS) vs $L_g$, (3) Energy-delay product ($E.\tau$) vs $L_g$ and
(4) $I_{on}/I_{off}$ vs $\tau$ are used to evaluate the performance
of the device. The simulated results are benchmarked against published state of
the art HEMT devices \cite{k07,k08} and planar/nonplanar MOS devices \cite{by02,c05,rc03}.
The proposed device shows good promise as one of the candidates to
replace existing MOS technology in the future.
\section{Transistor Design and Simulation}\label{sec:proposal}
The top view of the proposed device is shown
in Fig. \ref{fig:schematic}(a). The corresponding cross section along AA$^\prime$
is captured in Fig. \ref{fig:schematic}(b).
\bfg[htbp!]
\bc
\vs{-0.3in}
\includegraphics[scale=0.4,angle=0]{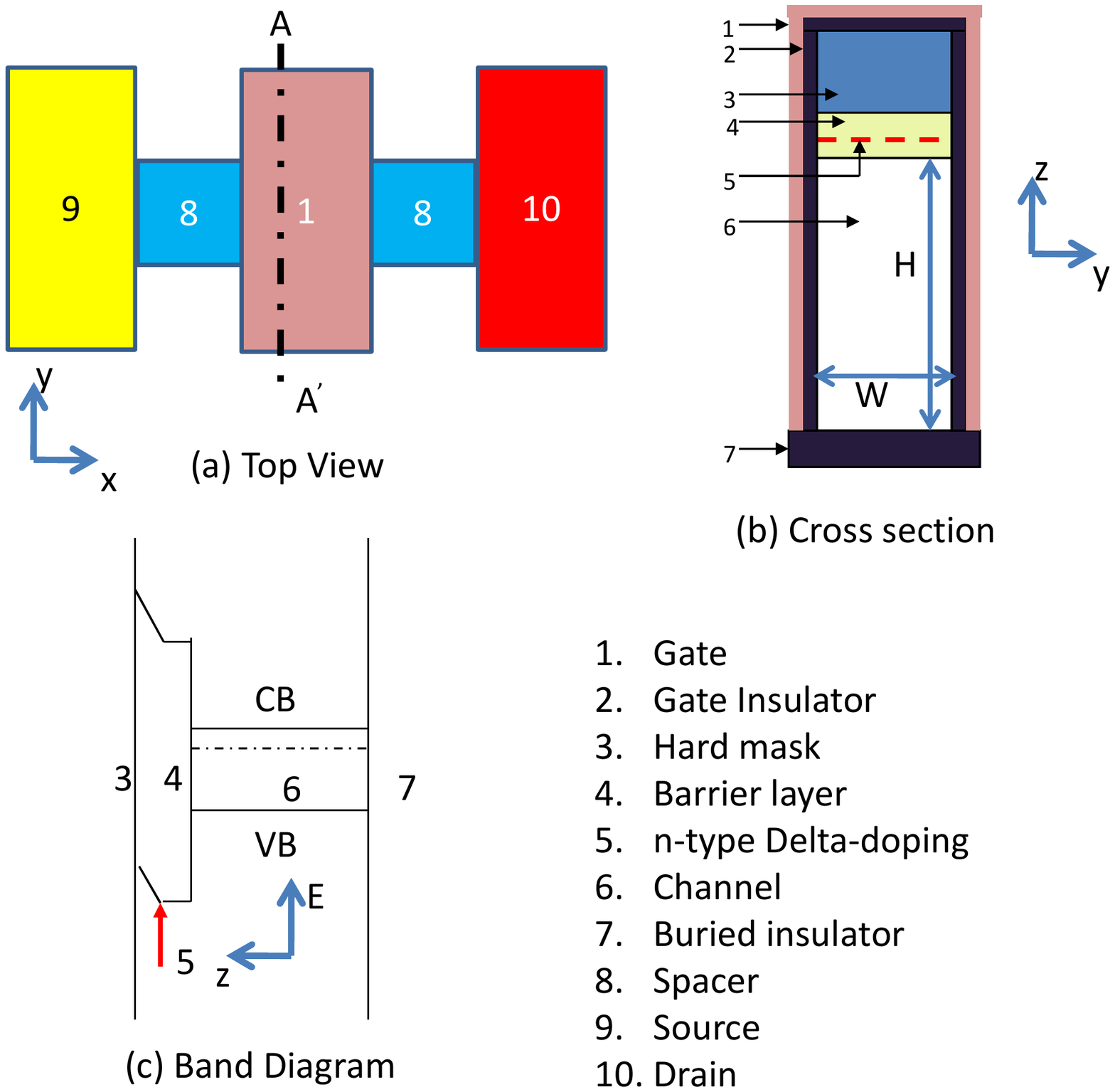}
\vs{-0.2in}
\caption{The (a) top view, (b) cross-section and (c) schematic band diagram in the $z$
direction of the proposed device HFinFET.}\label{fig:schematic}
\ec
\efg
The architecture resembles a FinFET with two vertical gates on the sides of
the rectangular In$_{0.53}$Ga$_{0.47}$As fin of width $W$ and height $H$
sitting on top of a burried insulator. However, instead of
the hard mask directly sitting on top of the channel, a barrier layer is
sandwiched between the channel and the hard mask. The barrier layer is
selected such that it has a conduction band edge offset with the channel.
Also, the barrier layer is almost lattice matched with the channel material
so as to minimize the traps at its interface with the channel.
The hard mask can be thick enough to make the channel immune to the
top gate to reduce so called corner effects.
The schematic band diagram in the vertical $z$ direction
is shown in Fig. \ref{fig:schematic}(c).
An n-type delta doped layer in the barrier layer can supply electrons in the
channel at ON state, like a HEMT operation. Using a mid-gap workfunction gate material
with no gate bias and n-doped source and drain,
the device conducts current, similar to a HEMT, providing excellent ON characteristics
due to bulk conduction. We also note that the small geometrical cross section of the fin and
the wavefunction pushing effect from the surface due to the
low effective mass of In$_{0.53}$Ga$_{0.47}$As help to
achieve volume inversion, causing a spread of carrier distribution all along the cross section
of the fin. Clearly, there is a distinction of turning ON
mechanism between the proposed
device and {\it`MOS-like'} devices including FinFET. The proposed device does
not require any gate induced surface inversion. This helps the transistor to
operate at very low electric field in the $yz$ plane achieving much lower carrier scattering rate.
However, unlike HEMT and more like a FinFET, the device is
turned off by applying negative bias at the side gates, thus achieving
excellent gate control providing good off state characteristics.

The device, as described in Fig. \ref{fig:schematic}, is operating in depletion mode where we need
to switch off the `normally ON' transistor by applying negative bias at the gates.
This depletion mode transistor can be converted into an enhancement mode
by using gate workfunction engineering \cite{k07,k08}.
In the rest of the paper, we assume a gate workfuntion such that the flatband voltage
is $V_{dd}$. It should be noted that this workfunction engineering
does not impact the low vertical field during ON state, but helps
to shift the threshold voltage of the device from negative to positive.
This flatband ON condition \cite{p06}, coupled with the bulk
conduction mechanism and inherently large
mean free path of In$_{0.53}$Ga$_{0.47}$As helps to operate
the transistor very close to the ballistic limit.

An effective mass ($m^*$) based 3D self-consistent Poisson-Schrodinger
solver coupled with ballistic transport model
has been developed to simulate the device.
Instead of bulk effective mass, we use a fin width dependent
in-plane and quantization effective mass as shown in
Fig. \ref{fig:fig2}(a). Since the fin height is comparatively
large, the effective mass is assumed to depend
only on fin width.
The in-plane effective mass $m_{zx}^*$ of In$_{0.53}$Ga$_{0.47}$As fin
is extracted by
separately fitting a parabola at the conduction band minimum
of the bandstructure (obtained from a sp$^3$d$^5$s$^*$
tight-binding method \cite{tbb04}) of a thin film of InAs and GaAs followed
by subsequent interpolation using Vegard law \cite{p08}.
The quantization mass $m_y^*$ is extracted by finding the energy difference of the
conduction band minimum between the bulk and the thin film followed by similar interpolation.
We also note that due to strong quantization and low voltage operation,
the satellite valley spill over effect is negligible in thin In$_{0.53}$Ga$_{0.47}$As channel and hence
all the carriers can be safely assumed to be in the $\Gamma$ valley \cite{c07}.
The ballistic transport model is one similar to the model in \cite{kn94} and does not take
into account the tunneling components of the drain current.
We analyze three different devices with $L_g$ as 20nm, 15nm and 10nm.
The doping density of the delta doped layer is assumed to be $2\times10^{12}$ cm$^{-2}$.
A gate insulator EOT of 1nm, fin height($H$) of 15nm and a gate to source/drain underlap
of 5 nm are assumed for all the transistors.
The hard mask thickness is assumed to be large enough so that the top gate does
not have any effect on the transistor electrostatics.
The fin widths ($W$) of different gate length devices are varied as 5nm, 7.5nm and 10nm.
The supply voltage has been taken to be 0.5V.
\section{Results and Performance Evaluation}\label{sec:resdults}
Fig. \ref{fig:fig2}(b) and (c) show typical output and transfer characteristics of the
device with $L_g$=20nm and $W$=10nm. We note the excellent drain current
saturation in Fig. \ref{fig:fig2}(b)
as compared to HEMT devices (\cite{k07,k08}). This, we believe, is arising because the proposed
device has significantly lower EOT and double gate structure and thus has a much better
gate control over the channel. For a given $V_g$, at relatively
large drain bias, the sensitivity of the
height of the source-channel barrier on the drain voltage is much less resulting in improved
drain current saturation.
\bfg[htbp!]
\bc
\vs{-0.2in}
\includegraphics[scale=0.35, angle=0]{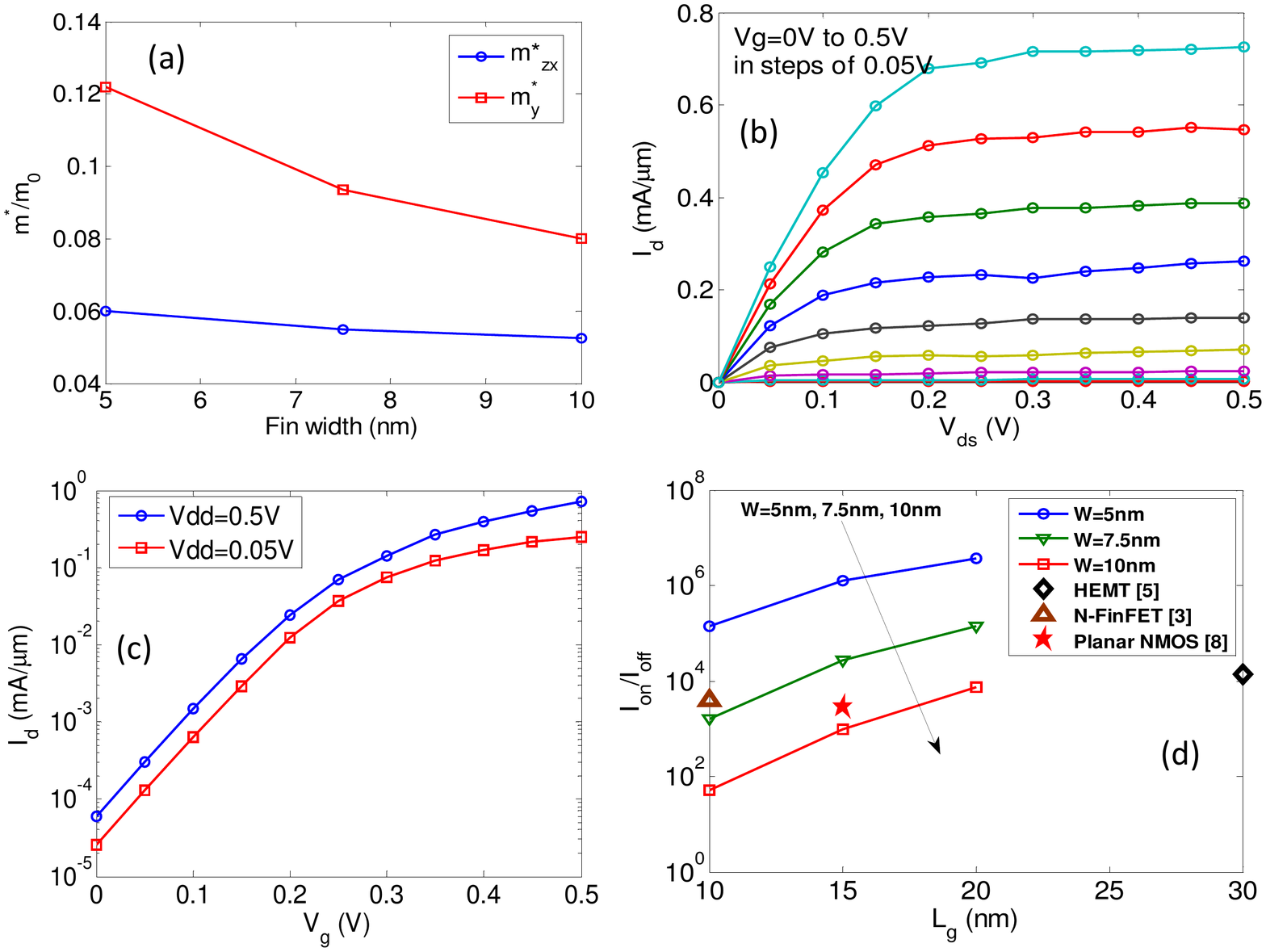}
\vs{-0.2in}
\caption{(a): The in-plane ($m_{zx}^*$) and quantization effective mass ($m_y^*$) of In$_{0.53}$Ga$_{0.47}$As
extracted from sp$^3$d$^5$s$^*$ tight binding method using Vegard law. Both of them decrease with
increase in fin width. (b): Simulated $I_d$-$V_{ds}$ characteristics of HFinFET for fin width of 10nm and $L_g$=20nm
shows excellent drain current saturation. (c): $I_d$-$V_g$ characteristics of HFinFET for fin width of 10nm and
$L_g$=20nm. (d): $I_{on}/I_{off}$ ratio of HFinFET with Vdd=0.5V for different Fin widths,
$R_s$=$R_d$=200$\Omega$-$\mu$m, compared against published HEMT, FinFET and Planar NMOS data.}\label{fig:fig2}
\ec
\efg
To evaluate the realistic ON state performance of the device, we assume source and drain series resistance
$R_s$=$R_d$=200$\Omega$-$\mu$m and their effects on transistor characteristics are taken into
account as a posteriori effect as explained in \cite{lnkk08}. Fig. \ref{fig:fig2}(d) plots
the $I_{on}/I_{off}$ as a function of $L_g$ for different $W$. Stronger quantization at
lower fin width reduces the OFF state leakage significantly and hence we observe improvement
in $I_{on}/I_{off}$ when we reduce the fin width. The $I_{on}/I_{off}$ ratio
of a 30nm gate length InAs HEMT, obtained
in \cite{k08} is plotted in the same figure for comparison.
Also, results for a 10nm gate length Si FinFET \cite{by02} and 15nm gate length planar
NMOSFET \cite{rc03} are compared in the same plot.

The proposed HFinFET has very impressive DIBL characteristics as shown in Fig. \ref{fig:fig3}(a)
For comparison, the DIBL values for published HEMT \cite{k08},
FinFET \cite{by02} and planar NMOS \cite{c05} are also
shown in the same plot. It is evident that HFinFET has excellent short channel behavior.
As a part of the benchmarking study of the device using the four performance
criteria described in \cite{c05}, we have plotted the subthreshold slope numbers,
obtained from the simulation, in Fig. \ref{fig:fig3}(b) and compared against different
state of the art devices.
\bfg[htbp!]
\bc
\vs{-0.3in}
\hs{-0.4in}
\includegraphics[scale=0.35, angle=0]{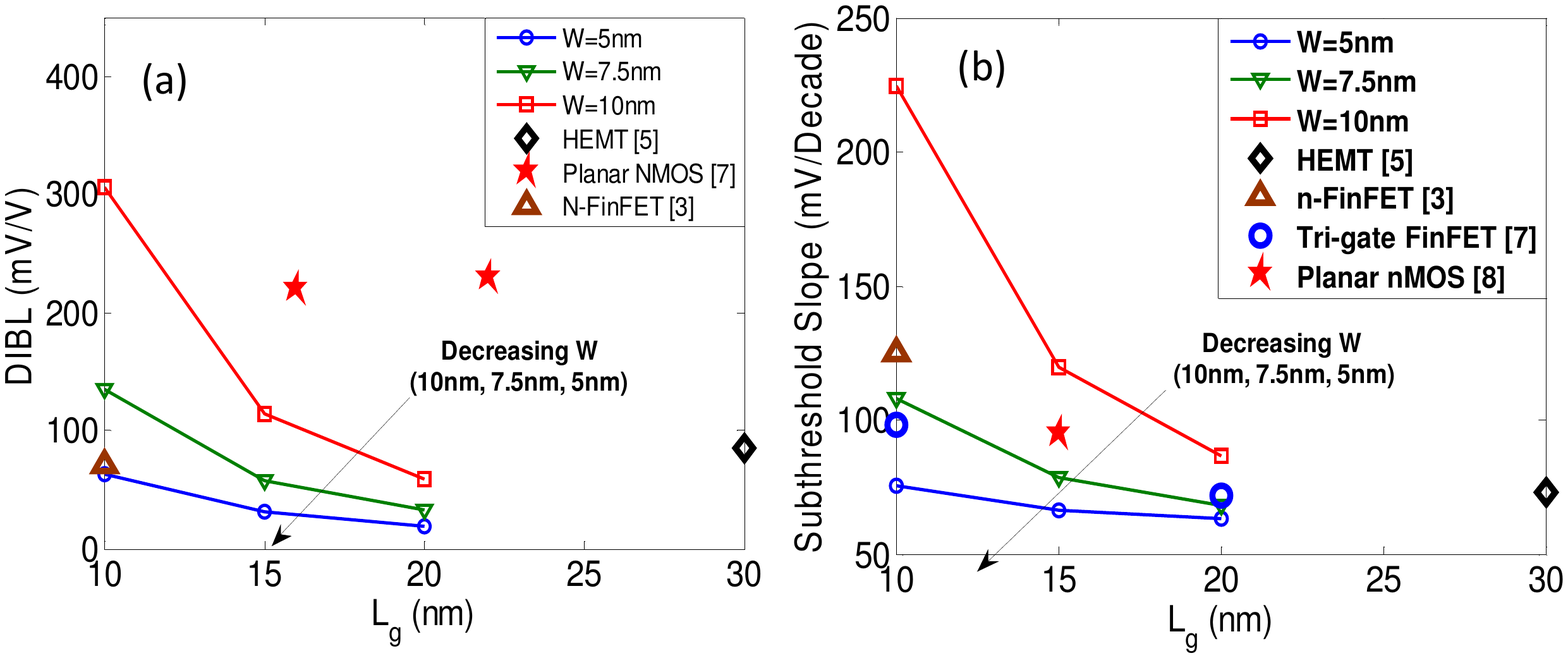}
\vs{-1.35in}
\caption{(a) DIBL and (b) Subthreshold slope characteristics of HFinFET,
compared against published HEMT, FinFET and Planar NMOS data.}\label{fig:fig3}
\ec
\efg
Fig. \ref{fig:fig4}(a) shows the intrinsic gate delay
as a function of $L_g$ for different fin widths. As expected, stronger quantization
arising from lower fin width degrades the ON current and hence gate delay. The
intrinsic gate delays obtained compare very well with published high performance HEMT
as well as planar and non-planar Si nMOSFETs, at different gate lengths.
Note that, at $L_g$=10nm, the HFinFET with $W$=7.5nm and $W$=10nm meets the
intrinsic gate delay projection of 0.167ps by ITRS 2007, however $W$=5nm misses the
target.
It is worthy to mention here that at smaller fin width, due to quantization,
the gate capacitance also reduces slightly, which actually compensates for the
delay degradation to some extent. Hence the ON current degradation is not completely
reflected in delay degradation. The switching of less charge due to quantization
impacts the energy-delay product of the device,
which is shown in Fig. \ref{fig:fig4}(b) and
compared against planar NMOSFET data from \cite{rc03}.
Lower energy-delay product of the transistor makes it suitable for low power
and high performance logic circuits.
\bfg[htbp!]
\bc
\vs{-0.3in}
\includegraphics[scale=0.35, angle=0]{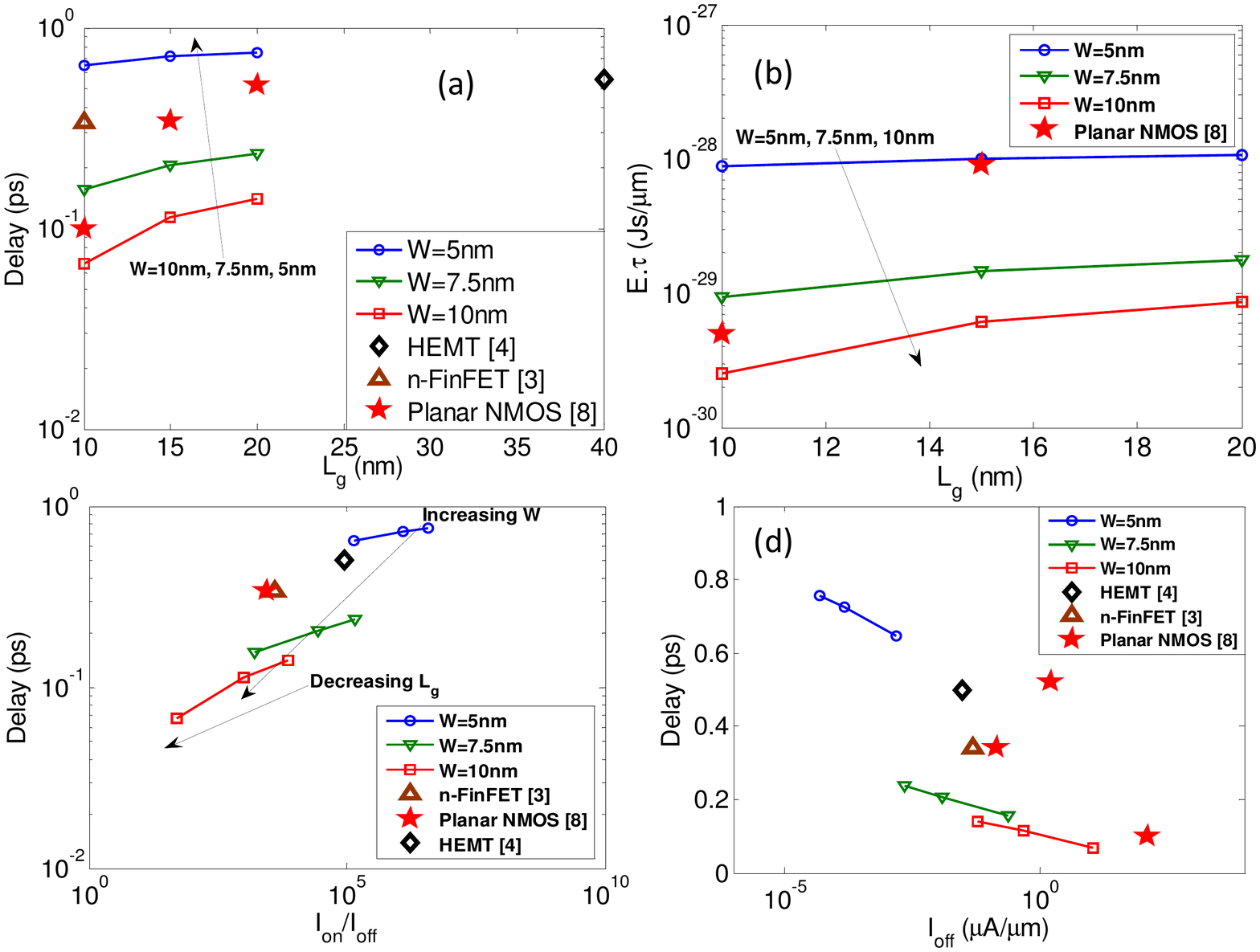}
\vs{-0.2in}
\caption{(a) Intrinsic gate delay (CV/I) and (b) Energy-delay product ($E.\tau$) of HFinFET
with Vdd=0.5V and $R_s$=$R_d$=200$\Omega$-$\mu$m.
Comparison of relative performance of HFinFET, HEMT, FinFET and Planar NMOS in the
(c) Delay versus $I_{on}/I_{off}$ design space and (d) Delay versus $I_{off}$ design space.
The relatively higher delay of the HEMT device is due to the larger channel length (40nm)
of the device.}\label{fig:fig4}
\ec
\efg
Finally, the simulated data points
are plotted in the delay versus $I_{on}/I_{off}$ design space in Fig. \ref{fig:fig4}(c) which
shows that narrower fin widths provide better ON to OFF current ratio, but
reducing the fin width too much kills the intrinsic gate delay.
When the results are compared against reported state of the art devices, HFinFET
is found to shift the operating points towards the ideal right bottom corner of the
design space.
To further probe the claim of better performance as well as lower OFF state current in HFinFET,
we have shown the device performance in the delay versus $I_{off}$ space in Fig. \ref{fig:fig4}(d).
Compared to existing devices, The proposed device shows
significantly improved
intrinsic gate delay at a given OFF state leakage showing
the potential to actually provide
high performance like HEMT/bulk nMOSFET and at the same time good
off state control like a FinFET.
\section{Conclusion}\label{sec:conclusion}
To conclude, in this letter, we have presented the design and
systematic simulation of a novel hybrid In$_{0.53}$Ga$_{0.47}$As n
channel transistor, called HFinFET, which uses the best of both the
worlds of a HEMT and a FinFET, i.e. high performance and good off
state control. The simulations results have been benchmarked using
standard device performance criteria and compared with available
state of the art devices including HEMT, bulk planar NMOS and
n-FinFET. Further investigation is required to optimize the
different device parameters. However, the preliminary study reveals
that an optimized HFinFET has the potential to be a promising
candidate for the next generation transistor to help continue
scaling.


\begin{thebibliography}{10}
\bibitem{ykc01}
Y. K. Choi, N. Lindert, P. Xuan, S. Tang, D. Ha, T. Anderson, S. J. King, J. Bokor and C. Hu,
``Sub-20nm CMOS FinFET Technologies," {\it IEDM Tech. Dig.}, pp. 412-424, 2001.
\bibitem{fly06}
F. L. Yang {\it et al.}, ``5nm-Gate Nanowire FinFET," {\it Symp. VLSI Tech.}, pp. 196-197, 2004.
\bibitem{by02}
B. Yu {\it et al.}, ``FinFET Scaling to 10nm Gate Length," {\it IEDM Tech. Dig.}, pp. 251-254, 2002.
\bibitem{k07}
D. H. Kim and J. A. Del Alamo, ``Logic Performance of 40nm InAs HEMTs," {\it IEDM Tech. Dig.}, pp. 629-633, 2007.
\bibitem{k08}
D. H. Kim and J. A. Del Alamo, ``30nm E-mode InAs PHEMTs for THz and Future Logic Applications,"
{\it IEDM Tech. Dig.}, pp. 2008.
\bibitem{d05}
S. Datta {\it et al.}, ``85nm Gate Length Enhancement and Depletion mode InSb Quantum Well Transistors
for Ultra High Speed and Very Low Power Digital Logic Applications," {\it IEDM Tech. Dig.}, pp. 763-766, 2005.
\bibitem{c05}
R. Chau, S. Datta, M. Doczy, B. Doyle, B. Jin, J. Kavalieros,
A. Majumdar, M. Metz, and M. Radosavljevic, ``Benchmarking Nanotechnology for High-Performance and Low-Power
Logic Transistor Applications," {\it IEEE Trans. Nanotech.}, Vol. 4, No. 2, pp. 153-158, 2005.
\bibitem{rc03}
R. Chau, B. Doyle, M. Doczy, S. Datta, S. Hareland, B. Jin, J. Kavalieros, and M. Metz,
``Silicon Nano-Transistors and Breaking the 10nm Physical Gate Length Barrier,"
{\it Dev. Res. Conf.}, pp. 123-126, 2003.
\bibitem{p06}
M. Passlack, K. Rajagopalan, J. Abrokwah, and R. Droopad, ``Implant-Free High-Mobility Flatband
MOSFET: Principles of Operation," {\it IEEE Trans. Elec. Dev.}, Vol. 53, No. 10, pp. 2454-2459, 2006.
\bibitem{tbb04}
T. B. Boykin, ``Valence band effective-mass expressions in the sp$^3$d$^5$s$^*$
empirical tight-binding model applied to a Si and Ge parametrization," {\it Phys. Rev. B}, 69, pp. 115201, 2004.
\bibitem{p08}
H. S. Pal T. Low and M. S. Lundstrom, ``NEGF Analysis of InGaAs Schottky Barrier Double Gate MOSFETs,"
{\it IEDM Tech. Dig.}, 2008.
\bibitem{c07}
K. D. Cantley Y. Liu, H. S. Pal, T. Low, S. S. Ahmed and M. S. Lundstrom,
``Performance Analysis of Ill-V Materials in a Double-Gate nano-MOSFET,"
{\it IEDM Tech. Dig.}, pp. 113-116, 2007.
\bibitem{kn94}
K. Natori, ``Ballistic metal-oxide-semiconductor field effect transistor,"
{\it J. Appl. Phys.}, Vol. 76, No. 8, pp. 4897, 1994.
\bibitem{lnkk08}
M. Luisier  N. Neophytou, N. Kharche, N. and G. Klimeck, ``Full-Band and Atomistic simulation
of Realistic 40nm InAs HEMT," {\it IEDM Tech. Dig.}, 2008.
\end{thebibliography}
\end{document}